\documentclass[journal]{IEEEtran}

\ifCLASSINFOpdf

\else

\fi

\usepackage{amsmath}
\usepackage{amsfonts}
\usepackage{amssymb}
\usepackage{hyperref}
\usepackage{cleveref}
\usepackage{cite}
\usepackage{xcolor}
\usepackage{graphicx}
\usepackage{subfigure} 
\usepackage{bm}

\DeclareMathOperator*{\argmin}{arg\,min}

\begin{document}

\title{Ambisonics Binaural Rendering via Masked Magnitude Least Squares}

\author{Or~Berebi,
        Fabian~Brinkmann, Stefan Weinzierl~\IEEEmembership{Senior Member,~IEEE},
        and Boaz~Rafaely,~\IEEEmembership{Senior Member,~IEEE}
\thanks{O. Berebi and B. Stefan Weinzierl are with the School of Electrical and Computer Engineering, Ben-Gurion University of the Negev, Beer-Sheva 84105, Israel.}
\thanks{F. Brinkmann and Stefan Weinzierl are with the Audio Communication Group, Technische Universi\"at Berlin, Einsteinufer 17c, Berlin, D-10587, Germany}}

\maketitle

\begin{abstract}
Ambisonics rendering has become an integral part of 3D audio for headphones. It works well with existing recording hardware, the processing cost is mostly independent of the number of sound sources, and it elegantly allows for rotating the scene and listener. One challenge in Ambisonics headphone rendering is to find a perceptually well behaved low-order representation of the Head-Related Transfer Functions (HRTFs) that are contained in the rendering pipe-line. Low-order rendering is of interest, when working with microphone arrays containing only a few sensors, or for reducing the bandwidth for signal transmission. Magnitude Least Squares rendering became the de facto standard for this, which discards high-frequency interaural phase information in favor of reducing magnitude errors. Building upon this idea, we suggest Masked Magnitude Least Squares, which optimized the Ambisonics coefficients with a neural network and employs a spatio-spectral weighting mask to control the accuracy of the magnitude reconstruction. In the tested case, the weighting mask helped to maintain high-frequency notches in the low-order HRTFs and improved the modeled median plane localization performance in comparison to MagLS, while only marginally affecting the overall accuracy of the magnitude reconstruction.
\end{abstract}


\IEEEpeerreviewmaketitle

\section{Introduction}
Head-Related Transfer Functions (HRTFs) contain auditory cues required for spatial hearing and are thus a fundamental ingredient for 3D audio reproduction via headphones (binaural rendering)~\cite{blauert1997spatial,Xie2013a}. For Ambisonics binaural rendering, the spatial resolution of the HRTFs is often limited, due to the finite number of sensors contained in practically available microphone arrays~\cite{Rafaely2019,zotter2019ambisonics}. This order limitation makes it challenging to find a perceptually well-behaved low-order HRTF representation, making the HRTF set a factor that limits the achievable quality of low-order Ambisonics binaural rendering.

The possibly most straight forward and na\"ive solution is to mathematically solve the problem in the least squares (LS) sense and will be introduced in Sec.~\ref{sec:background}. This, however, causes severe coloration and localization artifacts for low-order reproduction~\cite{Bernschtz2014, ben2017spectral, Ben-Hur2019b}. These artifacts can to some degree be mitigated by analytical filters to correct the diffuse-field coloration~\cite{ben2017spectral} and an order-dependent tapering (windowing) of the HRTF in the Ambisonics domain~\cite{Hold2019a} that can both be applied as post-processing steps. Alternatively, the HRTF can be pre-processed before transforming it into the Ambisonics domain. Applying the LS solution to HRTFs that share a common linear phase above a selected cut-off frequency can further improve the rendering quality~\cite{Rasumow2014,Zaunschirm2018}. This idea has inspired Magnitude Least Squares (MagLS) rendering~\cite{schorkhuber2018binaural}, which is the current de facto standard and is introduced in more detail in Section~\ref{sec:background}. It uses the LS solution below a cut-off frequency up to which it is physically correct, and ignores the phase above the cut-off frequency. This essentially optimizes the low-order HRTF for coloration, which is one, if not the most important audio quality~\cite{Rumsey2005}.

Despite the significant leap forward in audio quality brought about by MagLS, there is still room for improvement. Especially at low Ambisonics orders, HRTFs rendered with MagLS still suffer from poor median plane localization performance~\cite{Engel2022a}, as well as errors with respect to interaural level differences~\cite{berebi2023imagls}, interaural time differences, and diffuse-field coherence~\cite{Zaunschirm2018,zotter2019ambisonics}. To improve these aspects, we suggest an iterative optimization of the low-order Ambisonics HRTF representation by means of a neural network and perceptually motivated loss functions. In the following, we focus on a general introduction of the method and the resulting improvement of the modeled median plane localization performance.
\section{Mathematical Background}\label{sec:background}
This section provides a brief introduction to Ambisonics binaural reproduction, spherical harmonics transform, and state-of-the-art HRTF preprocessing for low-order Ambisonics binaural reproduction, which will serve as the benchmark for this study.

\subsection{Ambisonics Binaural Reproduction}
Binaural audio can be rendered by filtering the plane-wave density function of the sound field with the listener's HRTF. Alternatively, this operation can be formulated in the spherical harmonics (SH) domain as~\cite{rafaely2010interaural}
\begin{equation}
    \mathbf{p} = [\mathbf{\tilde{a}}_{nm}]^H\mathbf{H}_{nm}
\end{equation}
where $\mathbf{H}_{nm} \in \mathbb{C}^{(N+1)^2 \times 2}$ is the $N$th order SH coefficients vector for the left and right ears. The $N$th order Ambisonics signal is denoted as $\mathbf{a}_{nm} \in \mathbb{C}^{(N+1)^2 \times 1}$ and modified to $\mathbf{\tilde{a}}_{nm}=(-1)^m [\mathbf{a}_{n(-m)}]^*$. This represents the SH coefficients of the captured or simulated plane-wave density function, with $(\cdot)^*$ denoting the complex conjugate operator and $(\cdot)^H$ the Hermitian operator. The resulting ear pressure signals $\mathbf{p} \in \mathbb{C}^{1 \times 2}$ can be transformed back to the time domain with an inverse Fourier transform and played back to the listener via headphones. For brevity, we will omit the frequency dependency throughout.

The quality of the reproduced binaural signal depends on the Ambisonics order $N$. This order is limited by the minimum of the available orders of $\mathbf{H}_{nm}$ and $\mathbf{a}_{nm}$. Typically, $N$ is equal to the SH order of $\mathbf{a}_{nm}$ since the sound field is assumed to be captured by a low-order spherical microphone array~\cite{zotter2019ambisonics}. Thus, the SH representation of the HRTF needs to be limited to match the low order of $\mathbf{a}_{nm}$. This limitation can degrade the resulting binaural signals in both spectral and spatial quality~\cite{ben2017spectral}.

\subsection{HRTF Preprocessing}\label{sec:pre_pro}
The low-order SH coefficients $\mathbf{H}_{nm}$ are usually calculated from a spatially dense sampled HRTF set. The goal is to find $\mathbf{H}_{nm} =\left[ \mathbf{h}^l_{nm} , \mathbf{h}^r_{nm}   \right]$ such that its inverse spherical harmonics transform (ISHT) will match the input HRTF set. This problem can be written as:
\begin{equation}\label{eq:arg_min_problem}
    \mathbf{h}_{nm} = \argmin_{\mathbf{\hat{h}}_{nm}} \sum_{\Omega} \mathcal{L}(\underbrace{\mathbf{\hat{h}}_{nm}^H\mathbf{y}(\Omega)}_{=\hat{h}(\Omega)\text{ (ISHT)}}, h(\Omega) ) 
\end{equation}
where $\mathcal{L}(.,.)$ is a loss function modeling dissimilarity, $h(\Omega) \in \mathbb{C}$ is the left or right ear HRTF at direction $\Omega \equiv (\theta, \phi) \in \mathcal{S}^2$, which denotes the spatial angle composed of $\theta \in [0, \pi]$ (the colatitude) and $\phi \in [0, 2\pi]$ (the azimuth angle). The vector $\mathbf{y}(\Omega) = [Y_i(\Omega)]_{i=1}^{(N+1)^2} \in \mathbb{C}^{(N+1)^2}$ denotes the spherical harmonics of order $n$ and degree $m$ at direction $\Omega$ with $i$ being a single index depending on the ordering convention of the spherical harmonics~\cite{Rafaely2019}. Notice that the $ ^{l/r} $ superscript was omitted for the sake of brevity, although the problem in Eq.~\eqref{eq:arg_min_problem} applies to both ears.

The choice of $\mathcal{L}$ directly impacts the resulting low-order HRTFs $\hat{h}(\Omega)$. Over recent years, different approaches have been suggested.
\subsubsection{Least Squares}
Setting the loss function to
\begin{equation}\label{eq:LS_loss}
    \mathcal{L}_{\text{LS}} = |\mathbf{\hat{h}}_{nm}^H\mathbf{y} - h|^2
\end{equation}
results in the Least Squares (LS) formulation of Eq.~\eqref{eq:arg_min_problem} with the  closed-form solution
\begin{equation}\label{eq:LS_solution}
    \mathbf{h}_{nm} = \mathbf{Y}^\dagger \mathbf{h}
\end{equation}
where $(\cdot)^\dagger$ is the pseudo-inverse, $\mathbf{h}\in \mathcal{C}^{Q \times 1}$ is the left or right HRTF vector for the $\Omega$ grid samples at $Q$ directions and $\mathbf{Y} = \left[\mathbf{y}(\Omega_1),\dots , \mathbf{y}(\Omega_Q)\right] \in \mathcal{C}^{Q \times (N+1)^2}$ is the SH matrix for the $Q$ directions. This is one solution for the spherical Fourier transform of order $N$. The LS approach matches the HRTFs in magnitude and phase and is physically accurate for $N\gtrsim35$, but leads to a poor match when using low orders~\cite{ben2019efficient}. Due to this, the LS solution is suggested to be used up to a cutoff frequency of:
\begin{equation}\label{eq:cutoff}
 f_\mathrm{c} = \frac{c}{2 \pi r} N 
\end{equation}
This relation results from the $N \le k r$ equation, providing a frequency and order-dependent limit below which the spherical Fourier transform is physically correct, with $k = 
\left(2 \pi f \right) /c$ being the wave number and $r, c$ the listener head radius and the speed of sound, respectively.

\subsubsection{Magnitude Least Squares}
Setting the loss function to
\begin{equation}\label{eq:MagLS_loss}
    \mathcal{L}_{\text{MagLS}} = |\,\, |\mathbf{\hat{h}}_{nm}^H\mathbf{y}| - |h|\,\,|^2
\end{equation}
results in the Magnitude Least Squares (MagLS) formulation of Eq.~\eqref{eq:arg_min_problem}~\cite{schorkhuber2018binaural}. Unlike the LS problem, this problem does not have a closed-form analytic solution due to its non-convex nature. In practice, an iterative combination of the reconstructed phase from the previous frequency with the HRTF magnitude of the current frequency became the de facto standard~\cite{zotter2019ambisonics}. This can be written starting from $f = f_\mathrm{c}$ as:
\begin{equation}\label{eq:MagLS_solution}
    \mathbf{h}_{nm}(f) = \mathbf{Y}^\dagger\left[ |\mathbf{h}|e^{i \angle \left( \mathbf{h}^H_{nm}(f_{-1})\mathbf{Y}   \right) }\right]
\end{equation}
with $i = \sqrt{-1}$, $\angle(\cdot)$ denoting the angle, $| (\cdot)|$ denoting element wise absolute value, and $f_{-1}$ the previous frequency bin. 

Matching only magnitude for $f > f_\mathrm{c}$ with a fade between (\ref{eq:LS_solution}) and (\ref{eq:MagLS_solution}) below $f_\mathrm{c}$ has been shown to significantly improve the quality of low-order signals compared to the LS solution~\cite{zotter2019ambisonics}, making it the state-of-the-art choice for low-order rendering.

\begin{figure}
 \centerline{
 \includegraphics[width=0.42\textwidth,trim={10pt 12pt 25pt 11pt},clip]{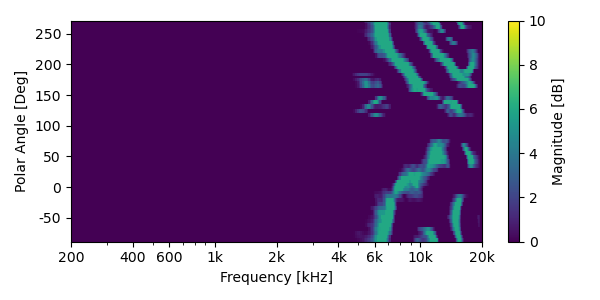}}
 \caption{Notch detection mask calculated for the right ear of KU100 HRTF with $k=4$ over the median plane.}
 \label{fig:notch_mask}
\end{figure}

\section{Proposed Method}
The LS and MagLS solutions illustrate that the loss function significantly impacts the binaural signal quality. Therefore, modifying it could potentially improve the performance considering human auditory perception. In this section, we introduce a modified loss function along with an optimization framework that aim at improving median plane localization for low-order Ambisonics binaural rendering.

\subsection{Masked MagLS}\label{sec:MMaglsLS}
The proposed loss named Masked MagLS (MMagLS) is defined as
\begin{equation}
    \mathcal{L}_{\text{MMagLS}} = |\,\, \left(|\mathbf{\hat{h}}_{nm}^H\mathbf{y}| - |h|\right)M\,\,|^2
\end{equation}
where $M(\Omega,f) \in \mathbb{R}$ is a spectro-spatial mask that increases the emphasis of the loss function at desired frequencies and directions. We chose $M$ as a notch mask to better preserve spectral notches in $\hat{h}(\Omega)=\mathbf{\hat{h}}_{nm}^H\mathbf{y}$. This has the potential to improve median plane localization where pinna-related notch patterns are deemed to be important~\cite{blauert1997spatial, Baumgartner2014}.

\subsection{Notch Mask}
Given the measured left and right HRTF, the notch mask $M$ is calculated for each ear individually by measuring the energy difference between the measured response and its one-octave smoothed version~\cite{tylka2017generalized}. This operation can be expressed as 
\begin{equation}\label{eq:MMagLS_loss}
    M(\Omega,f) =  \frac{|h_{smooth}(\Omega,f)|^2}{|h(\Omega,f)|^2}
\end{equation}
Next, the largest positive $k$ peaks are detected, and a bandwidth is determined around each peak by the closest left and right frequency bins where $M(\Omega,f) = 1$. All other entries of $M(\Omega,f)$ are set to $1$. Finally, the mask values are clipped to $4$ resulting in $1\leq M(\Omega, f)\leq 4$. Additionally, since we are interested in pinna-related notches, $ M(\Omega,f) = 1 $ below $ 4$ kHz and above $ 20 $ kHz as well as setting $ M(\Omega,f) = 1 $ for the contralateral ear direction within a $ 60^\circ $ great circle distance. This approach for generating the notch mask is inspired by the \emph{compare and squeeze} loudspeaker equalization method~\cite{Muller1999}. Figure~\ref{fig:notch_mask} shows $M(\Omega,f)$ in the median plane for the right ear of measured KU100 dummy head HRTFs~\cite{Bernschutz2013}.

\begin{figure}
 \centerline{
 \includegraphics[width=0.42\textwidth,trim={5pt 5pt 5pt 7pt},clip]{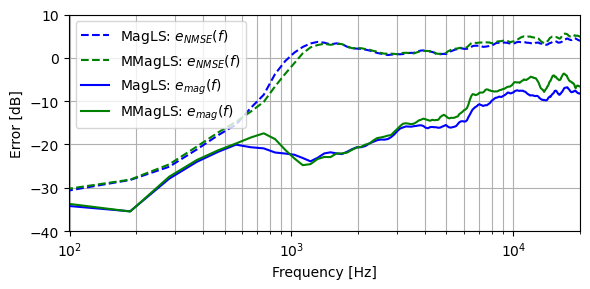}}
 \caption{Frequency errors of NMSE as in Eq.\eqref{eq:error_NMSE} and Magnitude as in Eq.\eqref{eq:error_mag}. Evaluated signals are $N=1$ MagLS and MMagLS, averaged over 2702 directions of a Lebedev grid}
 \label{fig:freq_errors}
\end{figure}

\begin{figure}
 \centerline{
 \includegraphics[width=0.42\textwidth,trim={0pt 0pt 0pt 0pt},clip]{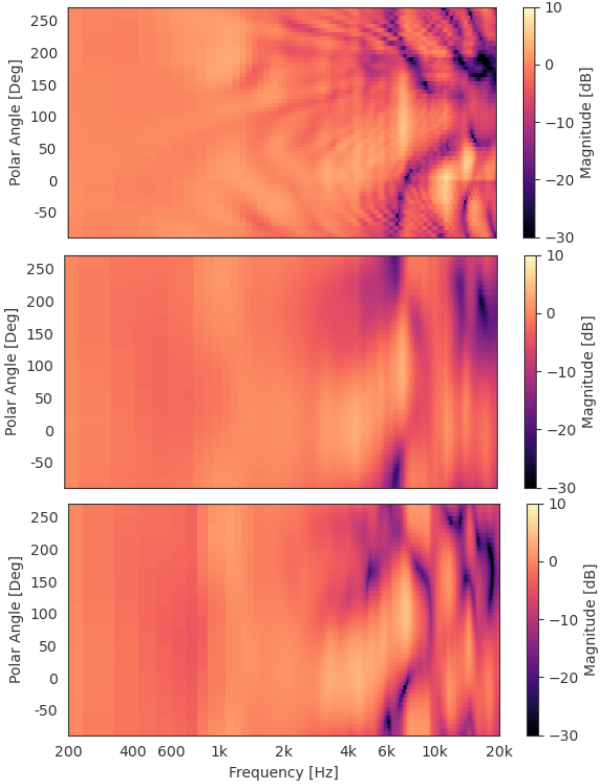}}
 \caption{Median plane spectra of KU100 for the reference HRTFs (top), $N=1$ MagLS (middle) and MMagLS (bottom).}
 \label{fig:spectra}
\end{figure}

\begin{figure}
 \centerline{
 \includegraphics[width=0.5\textwidth,trim={0pt 0pt 0pt 0pt},clip]{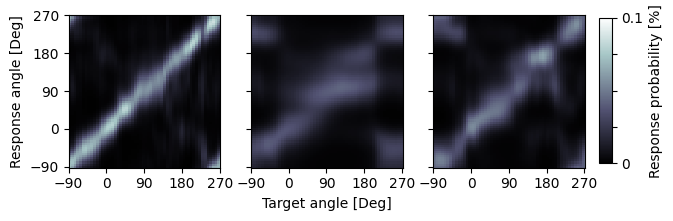}}
 \caption{Modeled median plane localization using the reference HRTFs as template and different target HRTFs. Left: reference HRTFs (QE=$14.0\%$, PE=$30.1^\circ$). Middle: MagLS (QE=$23.3\%$, PE=$45.1^\circ$). Right: MMagLS (QE=$17.8\%$, PE=$38.5^\circ$).}
 \label{fig:baumgartner2}
\end{figure}

\subsection{Neural Network Optimization}\label{sec:NN}
Solving Eq.~\eqref{eq:arg_min_problem} is performed using an optimization framework for HRTF SH preprocessing, which is based on solving optimization problems using deep neural networks~\cite{deng2014deep}. An initial low-order SH representation of the left and right HRTF is fed forward through a neural network. This can be written as: 
\begin{equation}\label{eq:NN}
    \mathbf{\hat{h}}_{nm} = f_{\bm{\beta}} \left( \mathbf{h}^{i}_{nm}  \right)
\end{equation}
where the neural network operation is denoted by $f_{\bm{\beta}}(\cdot)$, with $\bm{\beta}$ referring to the network's trainable parameters, and the initial SH representation is denoted by $\mathbf{h}^{i}_{nm}$. The low order HRTF $\hat{h}(\Omega,f)$ is computed from the resulting SH coefficients $\mathbf{\hat{h}}_{nm}$ by means of the ISHT, and $\mathcal{L}_{\text{MMagLS}}(\Omega,f)$~\eqref{eq:MMagLS_loss} and $\mathcal{L}_{\text{LS}}(\Omega,f)$~\eqref{eq:LS_loss} are computed with respect to the reference (input) HRTFs $h(\Omega,f)$.


The error is than converted to the scalar value
\begin{align}\label{eq:NN_error}
    \epsilon = \overline{\mathcal{L}_{\text{LS}}(\Omega, f)} +  \overline{\mathcal{L}_{\text{MMagLS}}(\Omega, f)}
\end{align}
with $\overline{\,\cdot\,}$ denoting the average across space and frequency.
Finally, the Adam optimizer is used to iteratively update $\bm{\beta}$ until $\epsilon$ reaches a local minimum to find the final MMagLS coefficients $\mathbf{\hat{h}}_{nm}$~\cite{kingma2014adam}.

\subsection{Error Measures}
The performance of $\mathbf{\hat{h}_{nm}}(f)$ is evaluated by comparing its ISFT to $h(\Omega,f)$. The first metric used is the normalized mean square error (NMSE), defined as:
\begin{equation}\label{eq:error_NMSE}
    e_{\text{NMSE}} (\Omega,f) = 10 \log_{10} \frac{|\mathbf{\hat{h}}_{nm}^H\mathbf{y} - h|^2}{|h|^2}
\end{equation}
The NMSE metric is sensitive to both magnitude and phase errors, making it a good measure of the physical accuracy of the binaural signals.

The second metric is the normalized magnitude error:
\begin{equation}\label{eq:error_mag}
    e_{\text{mag}} (\Omega,f) = 10 \log_{10} \frac{|\,\, |\mathbf{\hat{h}}_{nm}^H\mathbf{y}| - |h|\,\,|^2}{|h|^2}
\end{equation}
This metric is only sensitive to magnitude errors, which are the dominant cue for coloration and localization---at least at high frequencies~\cite{wightman1992dominant, Baumgartner2014}.


The final evaluation metric is a median plane localization model~\cite{Baumgartner2014} from the Auditory Modeling Toolbox v1.5~\cite{majdak_amt_2022}. The model compares reference and target HRTFs to compute a probabilistic matrix $P(\Omega)$ indicating the expected perceived source position under the assumption of familiarity with the reference $h$. The local polar error (PE), which reflects accuracy and precision for responses close to the target position, and the global performance quadrant error (QE), which relates to localization confusions, can both be calculated from $P(\Omega)$~\cite{middlebrooks1999virtual}. We assessed the baseline localization performance by providing the model with the reference HRTFs as target and template. The performance of MagLS and MMagLS was evaluated by using the reference HRTFs as template but the low-order HRTFs as target.

\begin{table}
\centering
\caption{Comparison of PE and QE increases for MagLS and MMagLS relative to the reference baseline error}
\label{table:1}
    \begin{tabular}{ |p{1.cm}|p{1.5cm}|p{1.5cm}|  }
         \hline
         \textbf{Method} & \textbf{PE increase}& \textbf{QE increase}\\ [0.5ex]
         \hline
         \hline
         MagLS & $15^\circ$ & $10\%$  \\
         \hline
         MMagLS & $\mathbf{8^\circ}$  & $\mathbf{4\%}$  \\
         \hline
        
    \end{tabular}
\end{table}

\section{Evaluation}
\subsection{Procedure}
KU100 HRTF~\cite{Bernschutz2013} was used to evaluate first-order Ambisonics ($N=1$) MMagLS to the state-of-the-art MagLS processing. The MagLS solution was computed using Eq.~\eqref{eq:LS_solution} below 680~Hz in accordance Eq.~\eqref{eq:cutoff} and Eq.~\eqref{eq:MagLS_solution} above. 
The MMagLS solution given by Eq.~\eqref{eq:NN_error} was calculated by evaluating $\mathcal{L}_{\text{LS}}$ in for frequencies $f < 680$ Hz and $\mathcal{L}_{\text{MMagLS}}$ above. Both terms in Eq.~\eqref{eq:NN_error} were evaluated for $2702$ directions in a Lebedev scheme. The MagLS coefficients were used as $\mathbf{h}^{i}_{nm}$ for neural network optimization. The neural network $f_{\bm{\beta}}(\cdot)$ used in the evaluation consisted of two linear layers operating on the input coefficients $\mathbf{h}^{i}_{nm}$ from the left and right, resulting in a relatively small $70k$ parameter network. Training, implemented using PyTorch, consisted of a 150-epoch loop, learning rate of $0.0005$, with $\mathbf{h}^{i}_{nm}$ represented by a complex value matrix of size of $\left[4, 257, 2\right]$ corresponding to $4$ SH coefficients, $257$ positive frequency bins and left and right ears. The training procedure taking about $2$ minutes to complete on a Apple M2 Max laptop.

\subsection{Results}
Figure~\ref{fig:freq_errors} presents $e_{\text{NMSE}}$ and $e_{\text{mag}}$ from Eqs.~\eqref{eq:error_NMSE} and~\eqref{eq:error_mag} averaged over $\Omega$. Both methods exhibit similar behavior, maintaining a low NMSE (less than $-20$~dB) up to the $f_\mathrm{c} = 680$ Hz cutoff frequency defined in Eq.~\eqref{eq:cutoff}, and high NMSE at 1~kHz and above. This similarity is expected as $\mathcal{L}_{\text{MMagLS}}$ does not affect phase, and the MagLS solution serves as its initial set.
Magnitude errors are also similar for both methods. Below the $f_\mathrm{c}$ , they employ an identical $\mathcal{L}_{\text{LS}}$ loss, and the loss is still similar for $f_\mathrm{c} < f < 6$~kHz due to the absence of pinna-related notch patterns. The average magnitude error between $6$ and $20$~kHz slightly increases from $-8.8$~dB for MagLS to $-6.1$~dB for MMagLS. This increase clearly stems from applying the notch mask, with MMagLS favoring magnitude accuracy at the regions of notches at the expense of decreasing the accuracy elsewhere.

Figure~\ref{fig:spectra} shows the HRTFs in the median plane. The distinct elevation-dependent notch patterns present in the reference  are clearly better preserved in the MMagLS solution than in the MagLS solution. The improvement of MMagLS over MagLS is also evident in the modeled localization performance in Fig.~\ref{fig:baumgartner2}. For the reference HRTFs, a clear and bright diagonal line emerges. This indicates a good localization accuracy where the virtual sound sources are perceived (response angle) close to or at the actual source position (target angle). For MagLS, the diagonal is blurred and additional bright spots occur at off-diagonal positions, indicating a poorer localization accuracy and an increased quadrant error rate. MMagLS, reduces the blur on the diagonal and off-diagonal bright spots, and can thus be expected to increase the localization accuracy with respect to MagLS. This is also reflected in the PE and QE. Compared to the reference, the QE increases by approx. $10$\% and the PE by $15^\circ$ for MagLS, whereas the increase in only about half as large for MMagLS as shown in table~\ref{table:1}. Finally, informal listening to a moving source auralization around the median plane suggests that MMagLS produces a narrower source width and more accurate localization compared to MagLS. However, these findings need to be verified through a formal listening test. Auralizations examples of MagLS and MMagLS renderings are available as supplemental material~\cite{berebi2024ambisonics}.
\section{Conclusions}
This paper introduces the masked MagLS method, which aims at improving the low-order HRTF representation for Ambisonics binaural rendering. We identified the MMagLS solution by minimizing loss functions by means of neural network optimization, and chose a mask targeted to preserve HRTF notches that are deemed crucial for accurate median plane localization. 

Evaluation results demonstrate the effectiveness of MMagLS and the neural network optimization framework in providing a more perceptually relevant low-order HRTF representation. The proposed method shows superior performance in maintaining spectral notches, resulting in an improved median plane localization performance on the localization model compared to the de facto standard MagLS rendering.

Future work will involve testing the proposed method on a broader set of HRTFs, conducting listening tests, introducing additional perceptually inspired loss terms, and evaluating various neural network architectures to further enhance the performance and applicability of the MMagLS method. Additional loss terms may well be combined in dependency of frequency, source position, and the application for which the HRTFs are intended. In particular, we plan to include loss functions accounting for interaural time and level differences and the interaural diffuse-field coherence, as recommended for first order rendering~\cite[Chap.~4.11.3]{zotter2019ambisonics}.

\ifCLASSOPTIONcaptionsoff
  \newpage
\fi

\newpage  
\newpage 

\bibliographystyle{./bibtex/IEEEtran}
\bibliography{./bibtex/refs}

\begin{thebibliography}{10}
\providecommand{\url}[1]{#1}
\csname url@samestyle\endcsname
\providecommand{\newblock}{\relax}
\providecommand{\bibinfo}[2]{#2}
\providecommand{\BIBentrySTDinterwordspacing}{\spaceskip=0pt\relax}
\providecommand{\BIBentryALTinterwordstretchfactor}{4}
\providecommand{\BIBentryALTinterwordspacing}{\spaceskip=\fontdimen2\font plus
\BIBentryALTinterwordstretchfactor\fontdimen3\font minus \fontdimen4\font\relax}
\providecommand{\BIBforeignlanguage}[2]{{%
\expandafter\ifx\csname l@#1\endcsname\relax
\typeout{** WARNING: IEEEtran.bst: No hyphenation pattern has been}%
\typeout{** loaded for the language `#1'. Using the pattern for}%
\typeout{** the default language instead.}%
\else
\language=\csname l@#1\endcsname
\fi
#2}}
\providecommand{\BIBdecl}{\relax}
\BIBdecl

\bibitem{blauert1997spatial}
J.~Blauert, \emph{Spatial hearing: the psychophysics of human sound localization}.\hskip 1em plus 0.5em minus 0.4em\relax MIT press, 1997.

\bibitem{Xie2013a}
B.~Xie, \emph{Head-Related Transfer Function and Virtual Auditory Display}, 2nd~ed.\hskip 1em plus 0.5em minus 0.4em\relax Plantation, FL, USA: J. Ross Publishing, 2013.

\bibitem{Rafaely2019}
B.~Rafaely, \emph{Fundamentals of Spherical Array Processing}, 2nd~ed.\hskip 1em plus 0.5em minus 0.4em\relax Berlin, Heidelberg, Germany: Springer, 2019.

\bibitem{zotter2019ambisonics}
F.~Zotter and M.~Frank, \emph{Ambisonics: A practical 3D audio theory for recording, studio production, sound reinforcement, and virtual reality}.\hskip 1em plus 0.5em minus 0.4em\relax Springer Nature, 2019.

\bibitem{Bernschtz2014}
B.~Bernsch{\"u}tz, A.~V. Giner, C.~P{\"o}rschmann, and J.~Arend, ``Binaural reproduction of plane waves with reduced modal order,'' \emph{Acta Acust. united Ac.}, vol. 100, no.~5, pp. 972--983, 2014.

\bibitem{ben2017spectral}
Z.~Ben-Hur, F.~Brinkmann, J.~Sheaffer, S.~Weinzierl, and B.~Rafaely, ``Spectral equalization in binaural signals represented by order-truncated spherical harmonics,'' \emph{The Journal of the Acoustical Society of America}, vol. 141, no.~6, pp. 4087--4096, 2017.

\bibitem{Ben-Hur2019b}
Z.~{Ben-Hur}, D.~L. Alon, B.~Rafaely, and R.~Mehra, ``Loudness stability of binaural sound with spherical harmonic representation of sparse head-related transfer functions,'' \emph{EURASIP Journal on Audio, Speech, and Music Processing}, vol. 2019, no.~1, p.~5, Mar. 2019.

\bibitem{Hold2019a}
C.~Hold, H.~Gamper, V.~Pulkki, N.~Raghuvanshi, and I.~J. Tashev, ``Improving {{Binaural Ambisonics Decoding}} by {{Spherical Harmonics Domain Tapering}} and {{Coloration Compensation}},'' in \emph{{{IEEE Int}}. {{Conf}}. {{Acoustics}}, {{Speech}} and {{Signal Processing}} ({{ICASSP}})}, Brighton, UK, May 2019, pp. 261--265.

\bibitem{Rasumow2014}
E.~Rasumow, M.~Blau, M.~Hansen, S.~{van de Par}, S.~Doclo, V.~Mellert, and D.~P{\"u}schel, ``Smoothing individual head-related transfer functions in the frequency and spatial domains,'' \emph{J. Acoust. Soc. Am.}, vol. 135, no.~4, pp. 2012--2025, Apr. 2014.

\bibitem{Zaunschirm2018}
M.~Zaunschirm, C.~Sch{\"o}rkhuber, and R.~H{\"o}ldrich, ``Binaural rendering of {{Ambisonics}} signals by head-related impulse response time alignment and a diffuseness constraint,'' \emph{J. Acoust. Soc. Am.}, vol. 143, no.~6, pp. 3616--3627, Jun. 2018.

\bibitem{schorkhuber2018binaural}
C.~Sch{\"o}rkhuber, M.~Zaunschirm, and R.~H{\"o}ldrich, ``Binaural rendering of ambisonic signals via magnitude least squares,'' in \emph{Proceedings of the DAGA}, vol.~44, 2018, pp. 339--342.

\bibitem{Rumsey2005}
F.~Rumsey, S.~Zieli{\'n}ski, R.~Kassier, and S.~Bech, ``On the relative importance of spatial and timbral fidelities in judgments of degraded multichannel audio quality,'' \emph{J. Acoust. Soc. Am.}, vol. 118, no.~2, pp. 968--976, Aug. 2005.

\bibitem{Engel2022a}
I.~Engel, D.~F.~M. Goodman, and L.~Picinali, ``Assessing {{HRTF}} preprocessing methods for {{Ambisonics}} rendering through perceptual models,'' \emph{Acta Acust.}, vol.~6, p.~4, 2022.

\bibitem{berebi2023imagls}
O.~Berebi, Z.~Ben-Hur, D.~L. Alon, and B.~Rafaely, ``imagls: Interaural level difference with magnitude least-squares loss for optimized first-order head-related transfer function,'' in \emph{Proceedings of the 10th Convention of the European Acoustics Association. FA 2023}, 2023, pp. 631--634.

\bibitem{rafaely2010interaural}
B.~Rafaely and A.~Avni, ``Interaural cross correlation in a sound field represented by spherical harmonics,'' \emph{The Journal of the Acoustical Society of America}, vol. 127, no.~2, pp. 823--828, 2010.

\bibitem{ben2019efficient}
Z.~Ben-Hur, D.~L. Alon, R.~Mehra, and B.~Rafaely, ``Efficient representation and sparse sampling of head-related transfer functions using phase-correction based on ear alignment,'' \emph{IEEE/ACM Transactions on Audio, Speech, and Language Processing}, vol.~27, no.~12, pp. 2249--2262, 2019.

\bibitem{Baumgartner2014}
R.~Baumgartner, P.~Majdak, and B.~Laback, ``Modeling sound-source localization in sagittal planes for human listeners,'' \emph{J. Acoust. Soc. Am.}, vol. 136, no.~2, pp. 791--802, Aug. 2014.

\bibitem{tylka2017generalized}
J.~G. Tylka, B.~B. Boren, and E.~Y. Choueiri, ``A generalized method for fractional-octave smoothing of transfer functions that preserves log-frequency symmetry,'' \emph{Journal of the Audio Engineering Society}, vol.~65, no.~3, pp. 239--245, 2017.

\bibitem{Muller1999}
S.~M{\"u}ller, ``Digitale {{Signalverarbeitung}} f{\"u}r {{Lautsprecher}},'' Ph.D. dissertation, RWTH Aachen, Germany, 1999.

\bibitem{Bernschutz2013}
B.~Bernsch{\"u}tz, ``A spherical far field {{HRIR}}/{{HRTF}} compilation of the {{Neumann KU}} 100,'' in \emph{{{AIA-DAGA}} 2013, {{International Conference}} on {{Acoustics}}}, Merano, Italy, Mar. 2013, pp. 592--595.

\bibitem{deng2014deep}
L.~Deng, D.~Yu \emph{et~al.}, ``Deep learning: methods and applications,'' \emph{Foundations and trends{\textregistered} in signal processing}, vol.~7, no. 3--4, pp. 197--387, 2014.

\bibitem{kingma2014adam}
D.~P. Kingma and J.~Ba, ``Adam: A method for stochastic optimization,'' \emph{arXiv preprint arXiv:1412.6980}, 2014.

\bibitem{wightman1992dominant}
F.~L. Wightman and D.~J. Kistler, ``{The dominant role of low-frequency interaural time differences in sound localization},'' \emph{The Journal of the Acoustical Society of America}, vol.~91, no.~3, pp. 1648--1661, 1992.

\bibitem{majdak_amt_2022}
\BIBentryALTinterwordspacing
{Majdak, Piotr}, {Hollomey, Clara}, and {Baumgartner, Robert}, ``Amt 1.x: A toolbox for reproducible research in auditory modeling,'' \emph{Acta Acust.}, vol.~6, p.~19, 2022. [Online]. Available: \url{https://doi.org/10.1051/aacus/2022011}
\BIBentrySTDinterwordspacing

\bibitem{middlebrooks1999virtual}
J.~C. Middlebrooks, ``Virtual localization improved by scaling nonindividualized external-ear transfer functions in frequency,'' \emph{The Journal of the Acoustical Society of America}, vol. 106, no.~3, pp. 1493--1510, 1999.

\bibitem{berebi2024ambisonics}
O.~Berebi, F.~Brinkmann, S.~Weinzierl, and B.~Rafaely, ``Ambisonics binaural rendering via masked magnitude least squares - supplemental material,'' \url{https://doi.org/10.5281/zenodo.14211191}, Nov. 2024.

\end{thebibliography}

\end{document}